\documentclass[letterpaper,twocolumn,10pt]{article}
\usepackage{usenix2019_v3}

\usepackage{amsmath,amssymb,amsfonts}
\usepackage{graphicx}
\usepackage{booktabs}
\usepackage{float}
\usepackage{listings}
\usepackage{subcaption}
\usepackage{xspace}
\usepackage{tikz}
\usetikzlibrary{arrows.meta, positioning, calc, shapes.geometric, fit, shadows}

\makeatletter
\def\@listi{\leftmargin\leftmargini
            \parsep 1\p@ \@plus 0.5\p@ \@minus 0.5\p@
            \topsep 2\p@ \@plus 1\p@ \@minus 1\p@
            \itemsep 1\p@ \@plus 0.5\p@ \@minus 0.5\p@}
\let\@listI\@listi
\makeatother

\makeatletter
\renewcommand\section{\@startsection {section}{1}{\z@}{-2.4ex plus -0.5ex minus -.2ex}{1.2ex plus .1ex}{\reset@font\large\bf}}
\renewcommand\subsection{\@startsection{subsection}{2}{\z@}{-2.0ex plus -0.5ex minus -.2ex}{0.9ex plus .1ex}{\reset@font\normalsize\bf}}
\makeatother

\let\oldthebibliography\thebibliography
\renewcommand{\thebibliography}[1]{%
  \oldthebibliography{#1}%
  \setlength{\itemsep}{0pt plus 0.3pt}%
  \setlength{\parsep}{0pt plus 0.3pt}%
}

\AtBeginDocument{\DeclareMathAlphabet{\mathcal}{OMS}{cmsy}{m}{n}}

\newcommand{\sab}{\mbox{SAB}\xspace}
\newcommand{\sqa}{\mbox{SQA}\xspace}
\newcommand{\openkedge}{\mbox{OpenKedge}\xspace}
\newcommand{\sal}{\mbox{SAL}\xspace}
\newcommand{\vai}{\mbox{VAI}\xspace}
\newcommand{\pdd}{\mbox{PDD}\xspace}

\definecolor{slate}{RGB}{112,128,144}
\definecolor{emerald}{RGB}{80,200,120}

\begin{document}

\title{\bf Sovereign Assurance Boundary: Certificate-Bound Admission for Agentic Infrastructure}

\author{
  {\rm Jun He}\\
  OpenKedge.io
  \and
  {\rm Deying Yu}\\
  OpenKedge.io
}

\maketitle

\begin{abstract}
Agentic infrastructure creates a control-plane authorization problem: non-deterministic reasoning systems may propose high-stakes mutations to production resources, but existing access controls (IAM), policy engines, consensus protocols, and audit logs either enforce static permissions or record actions only after execution. This paper introduces the Sovereign Assurance Boundary (\sab), a certificate-bound runtime admission boundary for autonomous execution authority in agentic infrastructure. \sab intercepts agent proposals at an \emph{assurance airlock}, compiles them into typed \emph{execution contracts} $C$, binds those contracts to cryptographic \emph{evidence digests} $H(E)$ and policy versions, and routes them through consequence-aware \emph{certification paths}. Successful admission emits a signed \emph{Sovereign Assurance Certificate} ($\Omega$) that is valid only for a scoped execution identity, revocation epoch, and validity window. A \emph{sovereign execution broker} verifies the certificate $\Omega$ and performs fresh pre-execution revocation and drift checks before invoking infrastructure APIs. We describe the airlock--broker architecture, formalize the admission and revocation invariants, and report preliminary feasibility measurements from a Go prototype over 2,500 admission attempts. The resulting model prevents autonomous reasoning alone from mutating state and turns delegated execution authority into a certificate-bound, evidence-bound, revocable, and replayable runtime artifact.
\end{abstract}

\section{Introduction}
\label{sec:intro}

Large language model (LLM) agents are increasingly connected to operational workflows rather than advisory interfaces, allowing them to scale resources, rotate credentials, deploy code, and export diagnostic data~\cite{yao-react, schick-toolformer}. In this setting, agentic infrastructure needs an admission mechanism that decides when a proposed action may become execution authority. Granting non-deterministic reasoning processes direct mutation rights over production control planes creates a safety and authorization risk: agents can become the proximate cause of physical or digital state transitions. Without a runtime admission boundary, individually plausible actions can accumulate into availability failures, data exposure, policy bypasses, or loss of forensic context.

\textbf{Sovereign Assurance Boundary (\sab)} interposes between agentic reasoning and high-stakes infrastructure mutation. We use ``sovereign'' operationally: the institution retains local control over policy evaluation, evidence retention, signing keys, revocation state, broker identity, and execution authority. It is not an agent monitor or a general governance process; it defines a certificate-bound admission boundary where an agent proposal is not treated as ambient authority. Under \sab, a proposal is eligible for execution only after it is compiled into a typed \emph{execution contract} $C$, bound to an \emph{evidence digest} $H(E)$ and policy version, assigned a consequence score $\mathcal{R}(C)$ and autonomy level $\mathcal{L}(C)$, routed through a consequence-aware \emph{certification path} $\operatorname{CertPath}(C)$, encoded as a signed certificate $\Omega$, checked against active revocation state, and verified by a \emph{sovereign execution broker} immediately before mutating the target system. The mechanism converts autonomous execution authority from an ambient permission into a revocable, evidence-bound runtime artifact.

\sab separates action-level and system-level assurance. Semantic Quorum Assurance (\sqa)~\cite{sqa-paper} serves as the action-level certification primitive. While \sqa validates individual proposed actions, \sab composes \sqa with evidence, policy, human approval, sovereign execution identity, revocation, and replay. \sab is independent of the SQA implementation: any validator mesh returning signed action-level attestations can instantiate $\Gamma_{\mathrm{SQA}}$. Action-level checks address unsafe individual mutations; system-level assurance binds execution to active policy, evidence, revocation state, and broker-enforcement paths.

\paragraph{Running Example.}
Consider an operations agent, $\mathcal{A}_{\text{ops}}$, tasked with resolving a service degradation. To restore reachability and diagnose the issue, $\mathcal{A}_{\text{ops}}$ proposes reconfiguring network security groups, scaling up compute nodes, rotating database credentials, rolling back a deployment, and exporting diagnostic logs. If an infrastructure stack lets that proposal directly invoke raw IAM permissions, autonomous execution of the sequence introduces systemic risk: the network change may expose private services, the resource expansion may exceed budget or quota constraints, credential rotation may trigger auth failures, the rollback may reintroduce vulnerabilities, and the log export may leak personally identifiable information (PII). The example shows why action-level checks alone do not bound institutional risk.

\paragraph{What SAB Is Not.}
\sab differs from existing controls in several ways:
\begin{itemize}
    \item \textbf{Not an IAM replacement:} IAM authorizes identities and requests; \sab certifies whether an autonomous proposal is semantically justified by evidence before a broker invokes those credentials.
    \item \textbf{Not a formal verification replacement:} \sab does not prove arbitrary safety, but composes formal proofs (when available) as inputs to its evidence digests.
    \item \textbf{Not a consensus protocol:} Consensus (e.g., Raft) agrees on state transitions; \sab evaluates whether a transition is institutionally authorized under current context.
    \item \textbf{Not only logging:} Logs record events post-execution; \sab binds pre-execution evidence and policy versions in the certificate $\Omega$, then links outcomes through a ledger-recorded outcome record $O$, making the admission process replayable.
    \item \textbf{Not only human-in-the-loop approval:} \sab is a system-level runtime admission boundary that cryptographically binds human signatures to specific contracts and evidence snapshots, preventing generic rubber-stamping.
\end{itemize}

We make four contributions:
\begin{enumerate}
    \item \textbf{Certificate-bound admission for autonomous infrastructure.}
    We identify a control-plane authorization gap that arises when
    non-deterministic agents are allowed to propose production
    mutations, and define the Sovereign Assurance Boundary (SAB) as
    a runtime admission boundary that converts agent proposals
    into certificate-bound execution authority.

    \item \textbf{A concrete airlock--broker architecture.}
    We introduce the assurance airlock, typed execution contract
    $C$, evidence digest $H(E)$, the certificate
    $\Omega$, and sovereign execution broker. Together, these
    mechanisms separate proposal, admission, and execution so that
    agent reasoning alone cannot directly mutate infrastructure state
    under the broker-enforced execution model.

    \item \textbf{Consequence-aware certification and revocation.}
    We formalize consequence scoring $\mathcal{R}(C)$, autonomy levels
    $\mathcal{L}(C)$, certification paths $\operatorname{CertPath}(C)$, monotone path
    selection, broker-side verification, and revocation epochs
    $\rho_{\mathrm{rev}}$. These mechanisms support escalation,
    rejection, suspension, and revocation before execution.

    \item \textbf{Prototype and feasibility evaluation.}
    We implement an initial Go prototype with contract serialization,
    OPA/Rego policy checks, Ed25519 certificate signing, a
    PostgreSQL-backed append-only ledger, revocation-epoch
    distribution, broker verification, and replay tooling. Using 500
    labeled execution contracts over 2,500 admission attempts, we
    report preliminary local feasibility measurements for admission
    latency, broker verification, revocation propagation, routing
    accuracy, replay completeness, and certificate overhead.
\end{enumerate}

\section{Background, Motivation, and System Requirements}
\label{sec:background-requirements}

\subsection{Background and Gap}
Securing agentic infrastructure requires integrating model-generated proposals with existing safety and security mechanisms.

\paragraph{Runtime Assurance.}
Traditional runtime assurance separates an advanced controller from a monitor or fallback controller to prevent safety envelope violations~\cite{simplex-architecture, bloem-shield-synthesis}. In cloud and database administration, mutations are discrete, policy-laden configuration changes. SAB adapts the runtime-assurance pattern by evaluating the institutional admissibility of proposals rather than continuous physical invariants.

\paragraph{Assurance Cases.}
Assurance cases present structured safety, security, or compliance arguments by linking claims to evidence~\cite{rushby-assurance}. They are usually static, design-time artifacts. SAB moves part of that structure into runtime admission by cryptographically binding telemetry, policies, and validator signatures to SAB certificates.

\paragraph{Access Control (IAM) and Policy Engines.}
Zero trust and identity management systems (IAM) verify caller identities and enforce structural policy compliance (e.g., via OPA Rego)~\cite{rose-zerotrust, aws-iam, opa-rego}. While IAM dictates whether an identity \emph{can} call an API, it cannot determine whether an autonomous proposal is \emph{semantically justified} by live evidence. SAB treats access control as a prerequisite, interposing a semantic certification boundary before credentials are used.

\paragraph{Consensus and Byzantine Fault Tolerance.}
Consensus protocols (e.g., Paxos, Raft, PBFT) replicate logs and enforce order across replicas~\cite{lamport-paxos, ongaro-raft}. They ensure agreement on state transitions but do not evaluate the semantic correctness or safety of those transitions. Semantic Quorum Assurance (SQA) certifies the semantic admissibility of individual proposed actions~\cite{sqa-paper}; SAB composes SQA into a system-level certification boundary.

\paragraph{AI Safety and Tool Governance.}
Agent safety tools (sandboxing, output filtering, model self-critiques) constrain agent execution environments~\cite{yao-react, owasp-llm-top10}. These mechanisms lack cryptographically signed, institutionally bound execution credentials or durable audit trails. SAB wraps these tools as untrusted validators and binds their collective output into SAB certificates.

\subsection{Motivation: Why Action Certification Is Not Enough}
Semantic Quorum Assurance (SQA) verifies locally whether an action is semantically admissible (e.g., whether a security-group change fits an active outage ticket). High-stakes infrastructure also needs a system-level check: whether the agent operates under active policies, fresh evidence, and current revocation rules.

A sequence of individually approved steps (e.g., opening a firewall, expanding an IAM role, rolling back a service, and exporting logs) can collectively create high-consequence availability, compliance, or security failures. Table~\ref{tab:sab-motivation} summarizes how SAB composes these security controls. SAB mediates the certificate-bound transition from proposal to execution, checking that the certificate $\Omega$ contains the evidence and validation epoch required by policy.

\begin{table*}[t]
\centering
\caption{Admission controls for agentic infrastructure. SAB certifies the institutional admission context surrounding individual actions.}
\label{tab:sab-motivation}
\scriptsize
\begin{tabular}{@{}p{0.13\linewidth}p{0.26\linewidth}p{0.18\linewidth}p{0.33\linewidth}@{}}
\toprule
\textbf{Control} & \textbf{Question} & \textbf{Artifact} & \textbf{Failure if Missing} \\
\midrule
Policy engine & Is the caller authorized for this operation? & Authorization token & Static permissions are treated as ambient authority. \\
SQA & Is the proposed action semantically admissible? & Action certificate & Locally unsafe mutations are executed. \\
Human approval & Does an operator approve this step? & Approval record & Approvals do not bind evidence or policy epochs. \\
Audit logging & What happened after execution? & Event log & Accountability is reconstructed post-hoc. \\
SAB & May this proposal become execution authority under institutional rules? & SAB certificate $\Omega$ & Individually safe actions cross into execution authority without system context. \\
\bottomrule
\end{tabular}
\end{table*}

\subsection{System Requirements}
To govern high-stakes agentic infrastructure, SAB must satisfy seven system requirements:
\begin{description}
    \item[\textbf{R1: Proposal/Execution Separation.}] Agent runtimes must have no standing credentials or direct execution privileges, treating them purely as action proposers.
    \item[\textbf{R2: Contractual Admission.}] Raw agent commands must compile into a strongly-typed execution contract $C$ defining target resources, preconditions, postconditions, and rollbacks.
    \item[\textbf{R3: Evidence-Bound Decisions.}] All admissions must bind to a cryptographic digest $H(E)$ of the telemetry and system state active at decision time.
    \item[\textbf{R4: Consequence-Aware Routing.}] Proposals must route to validation paths dynamically based on a consequence score $\mathcal{R}(C)$ (blast radius, irreversibility, data risk, and uncertainty).
    \item[\textbf{R5: Sovereign Execution.}] Admitted contracts must execute exclusively through an institutionally controlled broker that scopes credentials and validates the certificate $\Omega$.
    \item[\textbf{R6: Revocability.}] The system must support scoped withdrawal, enabling key rotation, validator quarantine, and domain suspension.
    \item[\textbf{R7: Replayability.}] SAB must record the certificate $\Omega$ and, after execution, outcome record $O$ in an append-only log, with the ledger binding $O$ to $\Omega$.
\end{description}

\section{SAB Architecture and Trust Model}
\label{sec:architecture}

The Sovereign Assurance Boundary (\sab) intercepts agent proposals before they can mutate production infrastructure. The system model separates unprivileged agent reasoning from privileged sovereign execution.

\subsection{System Model and Assumptions}
\label{sec:system-model}

We model the agent runtime as an untrusted proposer. It may be compromised, misaligned, or faulty, and has no standing administrative credentials. Downstream infrastructure APIs accept state mutations exclusively from the sovereign execution broker.

The trusted computing base (TCB) consists of the assurance airlock, policy engine, evidence ledger, revocation controller, private signing keys, and sovereign execution broker. The contract compiler, evidence collector, consequence scorer, SQA validator mesh, and approval console run outside the TCB but are security-relevant; Table~\ref{tab:sab-components} maps each component to its trust boundary and core systems function. The evidence collector is untrusted: while SAB detects unsigned or malformed evidence, signed false evidence from compromised authorized sources remains a residual risk. Similarly, the SQA mesh is an external, untrusted oracle: the airlock policy engine verifies validators' signed attestations $\Gamma_{\mathrm{SQA}}$ instead of trusting validator internal execution.

SAB assumes authenticated component identities, a versioned policy store, bounded clock skew, append-only ledger integrity, and timely propagation of revocation state. Let $pk_{\mathrm{SAB}}$ denote the airlock's public verification key and $\sigma_{\mathrm{SAB}}$ its signature on the certificate $\Omega$.

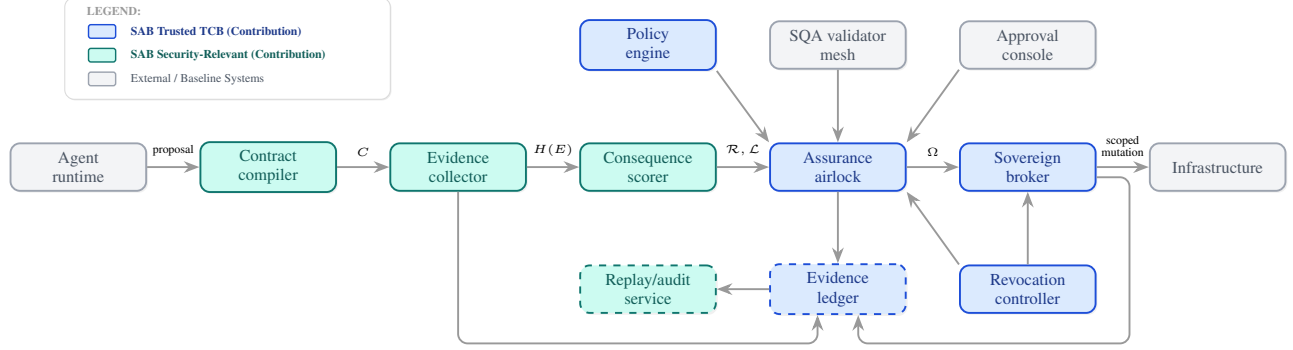
\begin{figure*}[t]
\centering
\definecolor{tcbFill}{RGB}{219, 234, 254}   
\definecolor{tcbBorder}{RGB}{29, 78, 216}   
\definecolor{tcbText}{RGB}{30, 58, 138}     

\definecolor{secFill}{RGB}{204, 251, 241}   
\definecolor{secBorder}{RGB}{15, 118, 110}  
\definecolor{secText}{RGB}{17, 94, 89}      

\definecolor{extFill}{RGB}{243, 244, 246}   
\definecolor{extBorder}{RGB}{156, 163, 175} 
\definecolor{extText}{RGB}{75, 85, 99}      

\resizebox{0.95\textwidth}{!}{%
\begin{tikzpicture}[
    >=Stealth,
    ext/.style={draw=extBorder, fill=extFill, text=extText, line width=0.8pt, rounded corners, align=center, minimum width=2.0cm, minimum height=0.70cm, font=\scriptsize, drop shadow={opacity=0.12, shadow xshift=0.8pt, shadow yshift=-0.8pt}},
    sec/.style={draw=secBorder, fill=secFill, text=secText, line width=0.8pt, rounded corners, align=center, minimum width=2.0cm, minimum height=0.70cm, font=\scriptsize, drop shadow={opacity=0.12, shadow xshift=0.8pt, shadow yshift=-0.8pt}},
    tcb/.style={draw=tcbBorder, fill=tcbFill, text=tcbText, line width=0.8pt, rounded corners, align=center, minimum width=2.0cm, minimum height=0.70cm, font=\scriptsize, drop shadow={opacity=0.12, shadow xshift=0.8pt, shadow yshift=-0.8pt}},
    secdashed/.style={draw=secBorder, fill=secFill, text=secText, line width=0.8pt, dashed, rounded corners, align=center, minimum width=2.0cm, minimum height=0.70cm, font=\scriptsize, drop shadow={opacity=0.12, shadow xshift=0.8pt, shadow yshift=-0.8pt}},
    tcbdashed/.style={draw=tcbBorder, fill=tcbFill, text=tcbText, line width=0.8pt, dashed, rounded corners, align=center, minimum width=2.0cm, minimum height=0.70cm, font=\scriptsize, drop shadow={opacity=0.12, shadow xshift=0.8pt, shadow yshift=-0.8pt}},
    conn/.style={draw=gray!80, ->, >=Stealth, line width=0.8pt}
]

\draw[draw=gray!30, fill=white, rounded corners, drop shadow={opacity=0.15, shadow xshift=0.5pt, shadow yshift=-0.5pt}] (-0.2, 1.0) rectangle (4.6, 2.5);
\node[anchor=west, font=\tiny\bfseries, text=gray!80] at (0.0, 2.3) {LEGEND:};

\draw[draw=tcbBorder, fill=tcbFill, rounded corners=1pt, line width=0.6pt] (0.15, 1.9) rectangle (0.55, 2.1);
\node[anchor=west, font=\tiny\bfseries, text=tcbText] at (0.65, 2.0) {SAB Trusted TCB (Contribution)};

\draw[draw=secBorder, fill=secFill, rounded corners=1pt, line width=0.6pt] (0.15, 1.55) rectangle (0.55, 1.75);
\node[anchor=west, font=\tiny\bfseries, text=secText] at (0.65, 1.65) {SAB Security-Relevant (Contribution)};

\draw[draw=extBorder, fill=extFill, rounded corners=1pt, line width=0.6pt] (0.15, 1.2) rectangle (0.55, 1.4);
\node[anchor=west, font=\tiny, text=extText] at (0.65, 1.3) {External / Baseline Systems};

\node[ext] (agent) at (0,0) {Agent\\runtime};
\node[sec] (compiler) at (2.8,0) {Contract\\compiler};
\node[sec] (evidence) at (5.6,0) {Evidence\\collector};
\node[sec] (score) at (8.4,0) {Consequence\\scorer};
\node[tcb] (airlock) at (11.2,0) {Assurance\\airlock};
\node[tcb] (broker) at (14.0,0) {Sovereign\\broker};
\node[ext] (infra) at (16.8,0) {Infrastructure};

\node[tcb] (policy) at (8.4, 1.8) {Policy\\engine};
\node[ext] (sqa) at (11.2, 1.8) {SQA validator\\mesh};
\node[ext] (human) at (14.0, 1.8) {Approval\\console};

\node[secdashed] (replay) at (8.4, -1.8) {Replay/audit\\service};
\node[tcbdashed] (ledger) at (11.2, -1.8) {Evidence\\ledger};
\node[tcb] (revoke) at (14.0, -1.8) {Revocation\\controller};

\draw[conn] (agent.east) -- node[above, font=\tiny, yshift=1pt] {proposal} (compiler.west);
\draw[conn] (compiler.east) -- node[above, font=\tiny, yshift=1pt] {$C$} (evidence.west);
\draw[conn] (evidence.east) -- node[above, font=\tiny, yshift=1pt] {$H(E)$} (score.west);
\draw[conn] (score.east) -- node[above, font=\tiny, yshift=1pt] {$\mathcal{R},\mathcal{L}$} (airlock.west);
\draw[conn] (airlock.east) -- node[above, font=\tiny, yshift=1pt] {$\Omega$} (broker.west);
\draw[conn] (broker.east) -- node[above, align=center, font=\tiny, yshift=1pt] {scoped\\[-1pt]mutation} (infra.west);

\draw[conn] (policy.south east) -- (airlock.north west);
\draw[conn] (sqa.south) -- (airlock.north);
\draw[conn] (human.south west) -- (airlock.north east);

\draw[conn] (revoke.north west) -- (airlock.south east);
\draw[conn] (revoke.north) -- (broker.south);

\draw[conn] (airlock.south) -- (ledger.north);

\draw[conn, rounded corners=4pt] (evidence.south) -- (5.6, -2.6) -- (10.9, -2.6) -- ([xshift=-0.3cm]ledger.south);

\draw[conn, rounded corners=4pt] ([yshift=-0.15cm]broker.east) -- (15.5, -0.15) -- (15.5, -2.6) -- (11.5, -2.6) -- ([xshift=0.3cm]ledger.south);

\draw[conn] (ledger.west) -- (replay.east);

\end{tikzpicture}}
\caption{Reference architecture for the Sovereign Assurance Boundary.}
\label{fig:sab-reference-architecture}
\end{figure*}

\begin{table*}[t]
\centering
\caption{SAB Component Reference.}
\label{tab:sab-components}
\scriptsize
\begin{tabular}{@{}p{0.17\linewidth}p{0.24\linewidth}p{0.53\linewidth}@{}}
\toprule
\textbf{Component} & \textbf{Trust Boundary} & \textbf{Core Systems Function} \\
\midrule
Agent Runtime & Untrusted / Outer & Proposes mutations as unprivileged JSON requests. \\
Contract Compiler & Security-relevant / Airlock & Normalizes natural-language or tool calls into structured contracts $C$. \\
Evidence Collector & Security-relevant / Airlock & Assembles telemetry, dependency logs, and tickets to compute evidence digest $H(E)$. \\
Consequence Scorer & Security-relevant / Airlock & Estimates blast radius and data risk; determines autonomy levels $\mathcal{L}(C)$. \\
Policy Engine & Trusted Control Plane & Enforces static rules and routes remaining ones to certification paths. \\
Assurance Airlock & Trusted Control Plane & Enforces path predicates, collects approvals, and issues signed SAB certificates $\Omega$. \\
SQA Validator Mesh & Fallible / Outer Mesh & Evaluates action semantics as decoupled, untrusted signing oracles. \\
Approval Console & Security-relevant / Outer & Presents structured contracts and evidence summaries for operator multi-signature. \\
Sovereign Broker & Trusted Control Plane & Verifies signatures and revocation status of $\Omega$; executes scoped actions through IAM. \\
Evidence Ledger & Trusted Control Plane & Durably records SAB certificates and bound outcome records $O$ in an append-only cryptographic log. \\
Revocation Controller & Trusted Control Plane & Monitors system state; advances revocation epochs $\rho_{\mathrm{rev}}$ and manages suspensions. \\
Replay Service & Security-relevant / Outer & Re-evaluates ledger records to reconstruct and audit historical admission decisions. \\
\bottomrule
\end{tabular}
\end{table*}

\subsection{SAB Invariant Properties}
\label{sec:sab-properties}

Under the trust model in Section~\ref{sec:system-model}, SAB enforces five authority-boundary invariants:

\begin{description}
    \item[\textbf{P1: Proposal-Execution Decoupling Invariant.}] Let $\mathrm{Mutate}(r, op)$ denote a state-changing operation on resource $r$. Under the assumption that the infrastructure API accepts mutations only from the sovereign broker, no agent proposal can directly trigger $\mathrm{Mutate}(r, op)$ without first being admitted by the airlock:
    \[
    \mathrm{Execute}(\Omega) \implies \mathrm{Admit}(C)
    \]
    where $C$ is the compiled execution contract corresponding to the proposal.

    \item[\textbf{P2: Certificate-Bound Execution Invariant.}] The sovereign execution broker executes a contract $C$ only after verifying the certificate $\Omega$ and its airlock signature:
    \[
    \mathrm{Execute}(\Omega) \implies \operatorname{Verify}(pk_{\mathrm{SAB}}, \Omega, \sigma_{\mathrm{SAB}}) \land C \in \Omega
    \]
    where the certificate is bound to the specific contract $C$, policy version $P_{\mathrm{ver}}$, execution identity $ID_{\mathrm{exec}}$, and validity window $T_{\mathrm{valid}}$.

    \item[\textbf{P3: Evidence-Binding Invariant.}] Approvals and validator attestations ($\Gamma_{\mathrm{SQA}}, \Gamma_{\mathrm{human}}$) are bound cryptographically to $C$ and the evidence digest $H(E)$:
    \[
    \operatorname{Approve}(C, H(E)) \implies \mathrm{Sign}(\Gamma, C \mathbin{\Vert} H(E))
    \]
    This binding prevents approval reuse across different contracts or stale environmental contexts.

    \item[\textbf{P4: Monotone Path Invariant.}] A contract $C$ is admitted only if all validation steps required by its selected certification path $\operatorname{CertPath}(C)$ are satisfied:
    \[
    \operatorname{Admit}(C) \implies \operatorname{VerifyPath}(C, \operatorname{CertPath}(C))
    \]
    where $\operatorname{CertPath}(C)$ depends on the consequence score $\mathcal{R}(C)$, autonomy level $\mathcal{L}(C)$, and policy version $P_{\mathrm{ver}}$. Policy may escalate a contract to stronger assurance or rejection, but lower-assurance paths cannot override requirements imposed by action class, consequence score, jurisdiction, or incident mode.

    \item[\textbf{P5: Ephemeral Authority Invariant.}] The sovereign execution broker rejects any certificate $\Omega$ if the current revocation epoch $\rho_{\mathrm{active}}$ is greater than the certificate's revocation epoch $\rho_{\mathrm{rev}}$, or if the execution time $t_{\mathrm{exec}}$ falls outside the validity window $T_{\mathrm{valid}}$:
    \[
    \mathrm{Execute}(\Omega) \implies \rho_{\mathrm{rev}} = \rho_{\mathrm{active}} \land t_{\mathrm{exec}} \in T_{\mathrm{valid}}
    \]
    The broker therefore rejects certificates from stale epochs.
\end{description}

\subsection{Proof Sketch}
\label{sec:proof-sketch}

\paragraph{Goal.} Establish that an untrusted agent runtime cannot directly execute a state-changing operation except through a broker-verified certificate satisfying the active policy path.

\paragraph{Assumptions.}
1. Infrastructure APIs accept mutations exclusively from the sovereign execution broker.
2. The broker verifies certificate signatures using the airlock public verification key $pk_{\mathrm{SAB}}$ and rejects invalid signatures.
3. Airlock private key remains uncompromised.
4. Active policy $P_{\mathrm{ver}}$ dictates path requirements.

\paragraph{Argument.}
To execute a mutation, the broker must verify the certificate $\Omega$ such that $C \in \Omega$ and $\operatorname{Verify}(pk_{\mathrm{SAB}}, \Omega, \sigma_{\mathrm{SAB}})$ (by P2). Because the airlock private key is secure, only the airlock could have produced $\sigma_{\mathrm{SAB}}$. The airlock emits this signature only if $\operatorname{Admit}(C)$ evaluates to true.
By P4, $\operatorname{Admit}(C)$ evaluates to true only if all path requirements $\operatorname{VerifyPath}(C, \operatorname{CertPath}(C))$ are satisfied under the policy $P_{\mathrm{ver}}$. Under the assumptions above, an agent proposal cannot execute unless it compiles to a contract $C$ satisfying the path checks mandated by institutional policy. Broker-enforced execution leaves agent reasoning with no direct mutation path (P1).

\subsection{Broker Verification Checks}
\label{sec:broker-checks}

Before invoking infrastructure APIs, the sovereign broker evaluates the pre-execution checks in Table~\ref{tab:broker-checks}. The checks cover certificate validity, contract identity, revocation state, policy epoch, state drift, and scoped credential issuance.

\begin{table}[h]
\centering
\caption{Broker verification checks.}
\label{tab:broker-checks}
\scriptsize
\begin{tabular}{@{}p{0.24\linewidth}p{0.28\linewidth}p{0.38\linewidth}@{}}
\toprule
\textbf{Broker Check} & \textbf{Input} & \textbf{Failure Behavior} \\
\midrule
Signature Verification & Certificate $\Omega$, public key $pk_{\mathrm{SAB}}$, signature $\sigma_{\mathrm{SAB}}$ & Reject execution, abort \\
Contract Match & Contract $C$, certificate contents $C \in \Omega$ & Reject execution, abort \\
Validity Window & Current time $t_{\mathrm{exec}}$, validity range $T_{\mathrm{valid}}$ & Reject execution, abort \\
Revocation Epoch & Certificate epoch $\rho_{\mathrm{rev}}$, active epoch $\rho_{\mathrm{active}}$ & Reject, fail-closed for $L_3/L_4$ \\
Policy Epoch & Certificate policy version $P_{\mathrm{ver}}$, active $P_{\mathrm{active}}$ & Reject execution, abort \\
State Drift Check & Bound evidence $E$, live state $S_{t_{\mathrm{exec}}}$, threshold & Reject, fail-closed for $L_3/L_4$ \\
Credential Issuance & Scoped identity $ID_{\mathrm{exec}}$, target resource API & Reject execution, abort \\
\bottomrule
\end{tabular}
\end{table}

\subsection{Worked Admission Example}

Consider an operations agent $\mathcal{A}_{\text{ops}}$ proposing an ingress network rule $C_{\mathrm{fw}}$ to resolve an outage. Table~\ref{tab:worked-example-stages} traces this proposal through the reference architecture.

\begin{table*}[t]
\centering
\caption{Worked example stages for security-group admission.}
\label{tab:worked-example-stages}
\scriptsize
\begin{tabular}{@{}p{0.09\linewidth}p{0.15\linewidth}p{0.20\linewidth}p{0.27\linewidth}p{0.19\linewidth}@{}}
\toprule
\textbf{Stage} & \textbf{Component} & \textbf{Output / Artifact} & \textbf{Example Value} & \textbf{Failure Check} \\
\midrule
1. Compile & Contract Compiler & Execution Contract $C_{\mathrm{fw}}$ & Add ingress rule; target: \texttt{sg-01}; port: 443; rollback: delete rule & Malformed syntax, unsupported API \\
2. Bind Evidence & Evidence Collector & Evidence Chain Digest $H(E_{\mathrm{fw}})$ & Telemetry (high latency); target config; incident ticket ID: \texttt{INC-101} & Stale telemetry, missing incident ticket \\
3. Score Consequence & Consequence Scorer & Consequence Score & $\mathcal{R}(C_{\mathrm{fw}}) = 8.5$; autonomy level $\mathcal{L}(C_{\mathrm{fw}}) = L_4$ & Score calculation failure \\
4. Route Path & Policy Engine & Certification Path & $\operatorname{CertPath}(C_{\mathrm{fw}}) = \mathrm{SQAPlusHuman}$ & Monotone path mismatch \\
5. Airlock Check & Assurance Airlock & SAB certificate $\Omega_{\mathrm{fw}}$ & Signed bundle including $C_{\mathrm{fw}}$, $H(E_{\mathrm{fw}})$, $\Gamma_{\mathrm{SQA}}$, $\Gamma_{\mathrm{human}}$ & Validator veto, missing human signature \\
6. Exec Check & Sovereign Broker & Broker execution decision & Verify $\sigma_{\mathrm{SAB}}$ with $pk_{\mathrm{SAB}}$, check $\Phi_{\mathrm{drift}}(E_{\mathrm{fw}}, S_{t_{\mathrm{exec}}})$, verify $\rho_{\mathrm{rev}}$ & Drift limit exceeded, stale epoch \\
7. Record Log & Evidence Ledger & Ledger transaction & Record $O_{\mathrm{fw}}$ bound to $\Omega_{\mathrm{fw}}$ & Ledger write failure \\
\bottomrule
\end{tabular}
\end{table*}

The example has two failure cases:
\begin{itemize}
    \item \textbf{Airlock policy rejection:} If $\mathcal{A}_{\text{ops}}$ attempts to open port 22 (SSH) to the public internet (\texttt{0.0.0.0/0}), the consequence scorer maps this high-risk action to $L_4$. The airlock's security-archetype validator detects a policy violation and triggers a critical veto, returning $\mathrm{Reject}$.
    \item \textbf{Broker-side verification failure:} If $C_{\mathrm{fw}}$ is admitted at time $t_1$, but before execution at $t_2$ the active revocation epoch advances from $\rho_1$ to $\rho_2$, the broker detects that $\rho_{\mathrm{rev}} = \rho_1 < \rho_2$ and aborts the execution. Similarly, if the broker's drift-check predicate $\Phi_{\mathrm{drift}}(E_{\mathrm{fw}}, S_{t_2})$ evaluates to false (indicating that the target security group's state has drifted since admission), the execution is aborted.
\end{itemize}

\section{Certificate-Bound Admission Pipeline}
\label{sec:admission-pipeline}

The admission pipeline evaluates proposals, scores consequence, routes them to certification paths, and issues signed certificates that the broker can verify before execution.

\subsection{Formal Admission Interface}
Let $C$ be an execution contract, $E$ be the evidence bound to that contract, $S_t$ be the observed infrastructure state at time $t$, $\mathcal{R}(C)$ be the consequence score, and $\mathcal{L}(C)$ be the assigned autonomy level; the certificate $\Omega$ is the resulting admission artifact. The airlock implements the admission function:
\[
    \operatorname{Airlock}(C, E, S_t) \rightarrow \Omega \;\; \text{or} \;\; \mathrm{Reject}.
\]
The function returns $\Omega$ only when the proposal satisfies the required certification path. Otherwise, it returns $\mathrm{Reject}$. To prevent time-of-check to time-of-use (TOCTOU) risks, the broker evaluates a state drift-check predicate $\Phi_{\mathrm{drift}}(E, S_{t_{\mathrm{exec}}})$ immediately before executing the mutation, verifying that the live state at execution time $S_{t_{\mathrm{exec}}}$ has not deviated from the bound evidence $E$ beyond policy safety margins.

\subsection{Consequence Scoring and Autonomy Levels}
The consequence score $\mathcal{R}(C)$ quantifies the operational, compliance, and security risk of the contract:
\begin{equation}
\label{eq:consequence-score}
\mathcal{R}(C) = \alpha B(C) + \beta P(C) + \gamma D(C) + \delta I(C) + \eta U(C).
\end{equation}
The terms measure specific risk vectors:
\begin{itemize}
    \item \textbf{Blast radius $B(C)$:} Affected resources, services, and network zones.
    \item \textbf{Privilege expansion $P(C)$:} Elevated credentials or access scopes required.
    \item \textbf{Data sensitivity $D(C)$:} Exposure risk of regulated data or PII.
    \item \textbf{Irreversibility $I(C)$:} Computational cost and difficulty of rolling back.
    \item \textbf{Uncertainty $U(C)$:} Lack of telemetry or confidence in evidence.
\end{itemize}

To tune the weights ($\alpha, \beta, \gamma, \delta, \eta$), security engineers run an offline \emph{adjudication loop} replaying historical incidents to tune classifications against expert consensus, biasing weights to escalate rather than under-classify when uncertainty $U(C)$ is high.

From the consequence score, active policy $\mathcal{P}_{\mathrm{inst}}$, and jurisdictional constraints $\mathcal{J}$, the airlock maps the contract to an autonomy level:
\[
\mathcal{L}(C) = \Lambda(\mathcal{R}(C), \mathcal{P}_{\mathrm{inst}}, \mathcal{J}).
\]
Table~\ref{tab:autonomy-levels} defines the consequence-aware autonomy levels.

\begin{table*}[t]
\centering
\caption{Consequence-aware autonomy levels.}
\label{tab:autonomy-levels}
\scriptsize
\begin{tabular}{@{}p{0.08\linewidth}p{0.21\linewidth}p{0.23\linewidth}p{0.40\linewidth}@{}}
\toprule
\textbf{Level} & \textbf{Class} & \textbf{Example Action} & \textbf{Assurance and Validation Obligation} \\
\midrule
$L_0$ & Advisory & Diagnostic query & Read-only; agent cannot compile or execute. \\
$L_1$ & Manual Draft & Batch restarts & Agent drafts contract; manual operator submission required. \\
$L_2$ & Bounded Autonomous & Stateless restart & Reversible, local action; OPA policy checks. \\
$L_3$ & Governed Autonomous & Autoscaling or rollback & Moderate blast radius; requires SQA and evidence binding. \\
$L_4$ & High-Stakes Autonomous & Security group, log export & High blast radius; SQA, human multi-sig, revocation checks. \\
$L_5$ & Prohibited & Deleting audit logs & Categorically denied autonomous execution. \\
\bottomrule
\end{tabular}
\end{table*}

\subsection{Certification Paths}
After assigning $\mathcal{L}(C)$, the airlock routes the contract through a certification path:
\begin{equation}
\label{eq:cert-path}
\operatorname{CertPath}(C) \rightarrow
\left\{
\begin{array}{@{}l@{}}
\mathrm{Reject},\ \mathrm{PolicyOnly},\\
\mathrm{LightweightQuorum},\ \mathrm{SQA},\\
\mathrm{SQAPlusHuman},\ \mathrm{Prohibited}
\end{array}
\right\}.
\end{equation}
The selected path determines the validation requirements:
\begin{itemize}
    \item \textbf{PolicyOnly:} Requires a valid contract, evidence digest, and a successful policy engine (e.g. OPA) check.
    \item \textbf{LightweightQuorum:} Requires multiple automated validator checks (syntax checks, telemetry checks, and dependency checks).
    \item \textbf{SQA:} Requires Semantic Quorum Assurance and validator diversity constraints.
    \item \textbf{SQAPlusHuman:} Requires SQA, critical-archetype veto checks, and operator multi-signature.
    \item \textbf{Prohibited / Reject:} Categorically disallowed or failed admission attempts.
\end{itemize}

\paragraph{Decoupled SQA Oracle.}
To keep the \sqa validator mesh outside \sab's trusted computing base, validator nodes run as external, isolated processes. They receive a read-only input bundle containing $C$ and $E$, and hold no credentials. Each validator $i$ independently evaluates the proposal and generates a cryptographically signed attestation:
\[
\Gamma_i = \mathrm{Sign}(\mathrm{key}_i, C \mathbin{\Vert} H(E) \mathbin{\Vert} \mathrm{vote}_i \mathbin{\Vert} \mathrm{metadata}_i).
\]
The airlock verifies validator signatures and checks diversity, quorum size, and veto requirements over the aggregated attestations $\Gamma_{\mathrm{SQA}} = \{ \Gamma_1, \Gamma_2, \dots \}$. Even under compromise, individual validators cannot forge approvals without private keys, nor can they directly mutate infrastructure.

\subsection{Admission Predicate}
To admit a contract, the airlock evaluates the admission predicate:
\begin{equation}
\label{eq:admission-predicate}
\begin{aligned}
\operatorname{Admit}(C) ={} &\Phi_{\mathrm{policy}}(C) \land \Phi_{\mathrm{evidence}}(C, E) \\
&\land \Phi_{\mathrm{assurance}}(C) \land \Phi_{\mathrm{authority}}(C).
\end{aligned}
\end{equation}
\begin{itemize}
    \item \textbf{$\Phi_{\mathrm{policy}}(C)$:} Policy permits the action class and target resource.
    \item \textbf{$\Phi_{\mathrm{evidence}}(C, E)$:} Required evidence exists, is fresh, and has trusted provenance.
    \item \textbf{$\Phi_{\mathrm{assurance}}(C)$:} The selected assurance path completed successfully.
    \item \textbf{$\Phi_{\mathrm{authority}}(C)$:} The broker identity, validity window, and revocation state are valid.
\end{itemize}

\subsection{SAB Certificates}
When the admission predicate is satisfied, the airlock issues the certificate $\Omega$:
\begin{equation}
\label{eq:assurance-certificate}
\begin{aligned}
\Omega = \langle
&C,\ H(E),\ \mathcal{R}(C),\ \mathcal{L}(C),\\
&\operatorname{CertPath}(C),\ \Gamma_{\mathrm{SQA}},\\
&\Gamma_{\mathrm{human}},\ ID_{\mathrm{exec}},\ T_{\mathrm{valid}},\\
&P_{\mathrm{ver}},\ \rho_{\mathrm{rev}},\ \sigma_{\mathrm{SAB}}
\rangle .
\end{aligned}
\end{equation}
where $ID_{\mathrm{exec}}$ is the execution identity, $T_{\mathrm{valid}}$ is the validity window, $P_{\mathrm{ver}}$ is the policy version, $\rho_{\mathrm{rev}}$ is the revocation epoch, and $\sigma_{\mathrm{SAB}}$ is the airlock signature. After execution, the broker emits a separate outcome record:
\begin{equation}
\label{eq:outcome-record}
O = \langle \Omega,\ \mathrm{status},\ t_{\mathrm{exec}},\ H(S_{\mathrm{post}}),\ \mathrm{result},\ \sigma_{\mathrm{broker}} \rangle ,
\end{equation}
where $\sigma_{\mathrm{broker}}$ is the broker signature. The ledger stores $O$ and binds it to $\Omega$.

\paragraph{Properties.}
\begin{itemize}
    \item \textbf{Replayability:} The replay service reconstructs the decision using $C$, $H(E)$, $\mathcal{R}(C)$, $\mathcal{L}(C)$, $\operatorname{CertPath}(C)$, $P_{\mathrm{ver}}$, and $\rho_{\mathrm{rev}}$.
    \item \textbf{Non-repudiation:} Approvals are cryptographically bound to $C$ and $H(E)$, preventing reuse or extraction.
    \item \textbf{Temporal Validity:} The certificate expires when the validity window $T_{\mathrm{valid}}$ closes or evidence freshness drops.
    \item \textbf{Outcome Binding:} Post-execution observations are bound to $\Omega$ through outcome record $O$, not embedded in the original certificate.
\end{itemize}
Table~\ref{tab:certificate-comparison} compares SAB certificates with standard change-auditing artifacts.

\begin{table*}[t]
\centering
\caption{Comparison of certification and auditing artifacts.}
\label{tab:certificate-comparison}
\scriptsize
\begin{tabular}{@{}p{0.21\linewidth}p{0.37\linewidth}p{0.36\linewidth}@{}}
\toprule
\textbf{Artifact} & \textbf{Records} & \textbf{Limitation} \\
\midrule
Audit log & Executed event, actor, time & Lacks admission context. \\
Policy decision record & Policy engine query result & Omits semantic evidence and validator votes. \\
SQA quorum & Validator votes and reasoning & Lacks system-level context and revocation epoch. \\
SAB certificate & Full pre-execution admission state & Outcomes require ledger-bound records and replay overhead. \\
\bottomrule
\end{tabular}
\end{table*}

\section{Revocation, Suspension, and Incident Response}
\label{sec:revocation-suspension}

Certificate-bound authority must be withdrawable. An incident may require immediate suspension across an action class, domain, validator, or execution path, so revocation and suspension are part of the control plane rather than an external operations procedure.

\subsection{Revocation Triggers and Withdrawal Mechanisms}
The controller monitors signals from the ledger, validator mesh, policy engine, broker, and compliance tools. Triggers apply globally or to a scoped domain $d$ (action class, service, region), as shown in Table~\ref{tab:revocation-triggers}.

\begin{table*}[t]
\centering
\caption{System revocation triggers.}
\label{tab:revocation-triggers}
\scriptsize
\begin{tabular}{@{}p{0.19\linewidth}p{0.37\linewidth}p{0.38\linewidth}@{}}
\toprule
\textbf{Trigger Class} & \textbf{Primary Signal} & \textbf{Institutional Consequence} \\
\midrule
Validator Degradation & Validator agreement rate or accuracy drops below threshold & Downgrade autonomy level, quarantine validator \\
Evidence Outage/Staleness & Telemetry streams, incident tickets, or configs become stale/unavailable & Airlock rejects new proposals in affected domain \\
Policy Drift/Rollback & Updated policy version $P_{\mathrm{ver}}$ published or rolled back & Invalidate active certificates, re-evaluate paths \\
Environmental Outage & System monitor detects live intrusion or active severity-1 outage & Advance revocation epoch $\rho_{\mathrm{rev}}$, suspend domain \\
Compromised Identity & Suspected leak of signing keys or broker execution credential & Revoke pending certificates, rotate execution keys \\
Operational Anomaly & Operator triggers emergency hold or manually rolls back change & Immediate human-only fallback, cancel active certificates \\
\bottomrule
\end{tabular}
\end{table*}

When a trigger fires, the controller applies one or more scoped withdrawal mechanisms (Table~\ref{tab:withdrawal-mechanisms}). Withdrawal is enforced at admission (the airlock consults suspension state before emitting $\Omega$) and execution (the broker checks revocation state before execution).

\begin{table*}[t]
\centering
\caption{Authority withdrawal mechanisms.}
\label{tab:withdrawal-mechanisms}
\scriptsize
\begin{tabular}{@{}p{0.19\linewidth}p{0.21\linewidth}p{0.54\linewidth}@{}}
\toprule
\textbf{Mechanism} & \textbf{Enforcement Point} & \textbf{Operational Effect} \\
\midrule
Level Downgrade & Consequence Scorer & Caps maximum permitted autonomy level $\mathcal{L}(C)$, routing actions to human approval. \\
Domain Suspension & Assurance Airlock & Categorically rejects all future proposals matching the suspended domain $d$. \\
Certificate Invalidation & Sovereign Broker & Broker rejects execution of unexecuted certificates matching the revoked class. \\
Validator Quarantine & SQA Mesh & Removes compromised or stale validators from participating in quorum validation. \\
Identity Rotation & Sovereign Broker & Rotates execution credentials, invalidating certificates bound to stale keys. \\
Human-only Fallback & Assurance Airlock & Bypasses automated routing, requiring manual approval for all actions. \\
Replay Sweep & Audit Service & Re-evaluates recent certificates under updated policy to identify active safety violations. \\
\bottomrule
\end{tabular}
\end{table*}

If the broker cannot obtain fresh revocation state (e.g., due to a network partition), it fails closed, rejecting all high-consequence ($L_3, L_4$) executions. Revocation updates are signed, versioned, and distributed as monotonically increasing epochs. Deployments that cache revocation state must bound cache lifetime and record the revocation epoch used for every broker decision.

\subsection{Formal Semantics}
Let $d$ be a domain or action class. Suspension is a future-admission invariant:
\begin{equation}
\label{eq:suspension-invariant}
\operatorname{Suspended}(d) \implies \forall C \in d,\; \operatorname{Admit}(C) = \mathrm{False}.
\end{equation}
The invariant is evaluated before path selection, preventing suspended domains from being bypassed.

Certificate revocation is an execution invariant:
\begin{equation}
\label{eq:revocation-invariant}
\operatorname{Revoked}(\Omega) \implies \operatorname{Execute}(\Omega) = \mathrm{False}.
\end{equation}
Equation~\ref{eq:revocation-invariant} applies to pending or unexecuted certificates. Completed executions cannot be undone; instead, the ledger marks $\Omega$ for post-hoc response and compensating actions.

\subsection{Incident Response Workflow and Override Risk}
During an incident, the controller identifies the affected domain $d$, installs a suspension or downgrade rule, and forces new proposals to human-only escalation. The broker invalidates pending certificates in $d$, while the replay service sweeps recent executions to link outcomes back to context.

Emergency overrides that bypass SAB remove the evidence and identity constraints required for accountability. Emergency operations should instead pass through SAB under stricter constraints: narrow scopes, short-lived certificates, and mandatory post-hoc replay. If an override is abused, the controller revokes pending certificates, quarantines keys, and replays the executions under standard policy.

\section{Case Studies}
\label{sec:case-studies}

The following case studies trace SAB's certificate-bound admission boundary across four operational settings (Table~\ref{tab:case-study-summary}).

\begin{table*}[t]
\centering
\caption{Case-study summary.}
\label{tab:case-study-summary}
\scriptsize
\begin{tabular}{@{}p{0.17\linewidth}p{0.21\linewidth}p{0.19\linewidth}p{0.15\linewidth}p{0.18\linewidth}@{}}
\toprule
\textbf{Case} & \textbf{Proposed Action} & \textbf{Risk} & \textbf{Certification Path} & \textbf{Outcome} \\
\midrule
Enterprise Cloud Operations & Open security-group rule during outage & Public exposure, privilege expansion, incident masking & \textsc{SQAPlusHuman} or \textsc{Reject} & Vetoed, escalated, or executed through broker with $\Omega$ \\
Generated-Code Deployment & Deploy agent-generated patch & Security regression, failed rollout, irreversible change & \textsc{SQA} or \textsc{SQAPlusHuman} & Certified artifact deployed with rollback and outcome evidence \\
Sovereign AI Cloud & Execute plan proposed by external foundation models & Jurisdictional leakage, external execution authority & \textsc{SQAPlusHuman} under local policy & Local certificate $\Omega$ and sovereign execution identity required \\
Regulated Data Workflow & Export logs or datasets for diagnosis & PII exposure, retention violation, cross-boundary transfer & \textsc{SQAPlusHuman}, \textsc{Reject}, or \textsc{Prohibited} & Redacted export, human approval, or denial with replay record \\
\bottomrule
\end{tabular}
\end{table*}

\subsection{Case Study 1: Enterprise Cloud Operations}
\paragraph{Scenario and Path.}
An operations agent $\mathcal{A}_{\text{ops}}$ detects network latency and proposes adding an ingress rule in a security group to bypass a blocked gateway. The compiler normalizes this into contract $C_{\mathrm{fw}}$, which the evidence collector binds to service-health metrics, incident ticket ID, and security-group configurations. Because modifying firewall rules alters production network posture, the scorer assigns a high blast-radius score, mapping $C_{\mathrm{fw}}$ to autonomy level $L_4$. If the proposed source range is public (\texttt{0.0.0.0/0}), the policy engine routes the contract to \textsc{SQAPlusHuman}. SQA validators review the proposal; a security-archetype validator holds veto power. If approved, the airlock signs certificate $\Omega_{\mathrm{fw}}$. The broker verifies $\Omega_{\mathrm{fw}}$ and evaluates the state drift-check predicate $\Phi_{\mathrm{drift}}$ before executing the API call.

\paragraph{Failure Prevented and Artifact.}
This path is intended to stop a latency diagnosis from becoming an overbroad network exposure or an unevidenced API call. The ledger records the contract, evidence digest $H(E)$, validator signatures, and broker decision, allowing post-hoc verification that the change was semantically justified.

\subsection{Case Study 2: Generated-Code Deployment}
\paragraph{Scenario and Path.}
A software engineering agent generates a bug patch and proposes deploying it to a Kubernetes cluster. The proposal is compiled into contract $C$, which includes source diffs, target service, and rollback plans. The collector binds $C$ to build provenance, test results, and vulnerability scans. The scorer distinguishes low-risk canary rollouts (routed to \textsc{SQA}) from privileged code changes or database migrations (escalated to \textsc{SQAPlusHuman}). The airlock evaluates PDD (Protocol-Driven Development) invariants, verifying that the generated patch matches interface definitions. The broker deploys only certified manifests and monitors rollout health, recording outcomes.

\paragraph{Failure Prevented and Artifact.}
The broker blocks unverified code from reaching production by rejecting proposals with missing rollbacks, failing tests, or unapproved dependencies. The certificate records all build and test evidence, enabling post-deployment replay.

\subsection{Case Study 3: Sovereign AI Cloud}
\paragraph{Scenario and Path.}
A sovereign operator uses a mixed model stack (external and domestic models) for planning. An external model proposes restarting services and exporting diagnostic telemetry. The local contract compiler decomposes the plan into separate contracts, and the collector binds local evidence (data-residency tags, configuration states). Sensitive context is held locally. The local policy engine enforces jurisdictional rules, rejecting contracts that violate residency policies or cross borders. Valid proposals are reviewed by local SQA validators, and high-stakes actions route to human operator multi-signature. The local broker executes approved contracts using domestic identities.

\paragraph{Failure Prevented and Artifact.}
The local broker path prevents external planning models from directly acquiring execution authority. External reasoning systems cannot obtain credentials, bypass jurisdictional policies, or execute mutations without local admission and replay. Certificate $\Omega$ records the domestic validation path and broker identity, making sovereign execution verifiable.

\subsection{Case Study 4: Regulated Data Workflow}
\paragraph{Scenario and Path.}
During an incident, an agent proposes exporting database logs for offline analysis. The logs may contain PII or regulated customer records. The compiler translates this into a data-export contract detailing source tables, destinations, and redaction rules. The collector binds schema metadata and legal-hold status. The scorer assigns a high data-sensitivity term, raising uncertainty if payloads are unclassified. If policy forbids autonomous export of the data class, the path is \textsc{Prohibited} or \textsc{Reject}. If export is permitted with redaction, the airlock enforces a minimization contract, routing the redacted export through compliance validators and human multi-signature.

\paragraph{Failure Prevented and Artifact.}
This path is intended to prevent autonomous leakage of regulated data under the guise of diagnostic utility. The airlock rejects overbroad exports or destinations lacking compliance controls, binding redaction attestations and human approvals to the export outcome.

\section{Evaluation Methodology and Targets}
\label{sec:evaluation}

We evaluate SAB's control-plane feasibility and performance tradeoffs with a Go prototype and three workload families.

\subsection{Prototype Implementation}
Our prototype of SAB in Go contains a stateless admission service, Protocol Buffer and JSON Schema contract definitions, an evidence-digest generator extracting Kubernetes and Active Directory logs, OPA/Rego policy evaluation, Ed25519 signing for $\Omega$, a PostgreSQL ledger, prototype target adapters, revocation epoch distribution, and a replay command-line interface. The target adapters simulate actions on Kubernetes and AWS/GCP APIs.

\subsection{Evaluation Methodology and Baselines}
We compare SAB against standard baseline security controls (Table~\ref{tab:baselines-comparison}) using 500 execution contracts: 200 cloud-infrastructure contracts (security groups, scaling, restarts, credential rotations), 150 CI/CD contracts (rollbacks, patch deployments, manifest updates), and 150 data-governance contracts (migrations, log exports, resource deletion).

\begin{table*}[t]
\centering
\caption{System baselines and comparison matrix.}
\label{tab:baselines-comparison}
\scriptsize
\begin{tabular}{@{}p{0.17\linewidth}p{0.27\linewidth}p{0.50\linewidth}@{}}
\toprule
\textbf{Baseline} & \textbf{Mechanism} & \textbf{Assurance Limitation} \\
\midrule
Static IAM/Policy & Checks static roles (e.g., AWS IAM, OPA) & Context-unaware; permits structurally valid but semantically unsafe actions. \\
Human Approval Only & Routes all proposals to human change queue & Introduces high latency; prone to rubber-stamping due to review fatigue. \\
Audit Logging Only & Records actions to security log after execution & No runtime prevention; cannot halt unauthorized actions before they occur. \\
LLM Self-Review & Proposing model critiques its own output & Susceptible to hallucinations and adversarial jailbreaks; no institutional control. \\
SQA without SAB & Quorum validation of individual proposed actions & Lacks system-level context: no evidence binding, revocation, or execution decoupling. \\
SAB (No Revocation) & Airlock admission without broker revocation checks & Vulnerable to stale certificates during outages or policy changes. \\
SAB (No Routing) & Fixed high-stakes certification path for all actions & High operational friction for low-risk actions; excessive latency. \\
\bottomrule
\end{tabular}
\end{table*}

\subsection{Experimental Setup}
Table~\ref{tab:preliminary-results} reports preliminary measurements from a single-node local workstation testbed, not a production cloud deployment. The topology contains one admission service, one policy engine, one PostgreSQL-backed ledger, one revocation service, three broker processes, three target-service adapters, and two workload-driver processes. The 500 contracts are independently labeled by three expert systems/security engineers for expected admission decision, certification path, and safety outcome, with disagreements resolved by majority vote and discussion. We ran five full-workload trials (2,500 admission attempts). Live local measurements include admission, OPA evaluation, Ed25519 signing, ledger writes, broker verification, revocation-epoch propagation, and SQA latency using a three-validator mesh with 2-of-3 quorum, Ed25519-signed attestations, and a security-archetype veto. Kubernetes, Active Directory, and cloud-control-plane evidence is replayed from captured or synthetic traces; target mutations are simulated by service adapters. Unless a row states otherwise, latency results are p50; SQA admission reports p50/p95, and rate metrics are percentages.

\subsection{Target Metrics and Expected Tradeoffs}
Table~\ref{tab:preliminary-results} reports the evaluation targets and measured results, grouped into latency overheads, routing accuracy, and storage metrics.

\begin{table*}[t]
\centering
\caption{Preliminary, workload-dependent evaluation results. Metric names specify p50, p95, mean, or percentage.}
\label{tab:preliminary-results}
\scriptsize
\begin{tabular}{@{}p{0.25\linewidth}p{0.19\linewidth}p{0.13\linewidth}p{0.35\linewidth}@{}}
\toprule
\textbf{Metric} & \textbf{Prototype Target} & \textbf{Measured Result} & \textbf{Expected Tradeoffs} \\
\midrule
PolicyOnly admission latency (p50) & $< 5$ ms & 1.84 ms & Fast path for low-consequence rules (OPA check). \\
SQA admission latency (p50/p95) & report p50/p95 by validator config & 185 ms / 380 ms & Parallel validator queries. \\
Broker verification latency (p50) & $< 1$ ms & 0.15 ms & Signature and active revocation epoch check. \\
Revocation propagation latency (p95) & $< 100$ ms & 12.4 ms & Delay to propagate new $\rho_{\mathrm{rev}}$ to local brokers. \\
Schema replay completeness (\%) & $100\%$ & $100\%$ & Fraction of ledger entries with replayable schema context. \\
Unsafe admission rate (\%) & lower is better & 0.4\% & Admissions that violate safety policy or human intent. \\
Routing accuracy (\%) & agreement with expert-labeled oracle & 96.8\% & Agreement of consequence scorer with expert-labeled paths. \\
Human approval reduction (\%) & $> 60\%$ & 68.2\% & Actions automated compared to a human-only queue. \\
Certificate storage overhead (mean) & $< 10$ KB / action & 3.4 KB & Space required on the evidence ledger per contract. \\
\bottomrule
\end{tabular}
\end{table*}

These measurements indicate control-plane feasibility on a single workstation, not production cloud performance. SAB trades latency for assurance, additional control-plane components for replayability, and occasional false escalation for lower under-classification risk. SQA and manual reviews add latency; evidence collection and ledger writes add storage and processing overhead while supporting audit and replay.

\section{Security Analysis}
\label{sec:security-analysis}

SAB mitigates authorization risks by decoupling proposal generation from execution through certificate-bound admission. The analysis defines the threat model, examines ten threat categories, and states the authority-boundary invariants.

\subsection{Threat Model and TCB Boundaries}
The threat model assumes that the agent runtime may be compromised (e.g., via prompt injection or credential theft) or faulty. Upstream evidence sources, SQA validators, and operators may also fail. We assume the integrity of the core control-plane components. Table~\ref{tab:tcb} maps the TCB boundaries.

\begin{table}[h]
\centering
\caption{Trusted and untrusted components in the SAB threat model.}
\label{tab:tcb}
\scriptsize
\begin{tabular}{@{}p{0.22\linewidth}p{0.24\linewidth}p{0.44\linewidth}@{}}
\toprule
\textbf{Component} & \textbf{TCB Status} & \textbf{Failure Mitigation} \\
\midrule
Agent runtime & Untrusted proposer & Airlock interposition, rate limiting \\
SQA validators & Untrusted external oracles & Diversity, vetoes, signing keys \\
Evidence collector & Outer engine: untrusted; signature verifier: trusted & Freshness, cryptographic provenance verification \\
Policy / revocation & Trusted control plane & Versioned policies, epoch gating \\
Airlock and broker & Trusted control plane & Hardened runtimes, key rotations \\
Evidence ledger & Trusted control plane & Cryptographic chains, backups \\
\bottomrule
\end{tabular}
\end{table}

\subsection{Threat Analysis}
Table~\ref{tab:security-threats} maps SAB's controls and residual risks across ten threat categories: agent compromise, validator failure, policy degradation, operational bypass, and authority creep.

\begin{table*}[t]
\centering
\caption{Threats, controls, and residual risks.}
\label{tab:security-threats}
\scriptsize
\begin{tabular}{@{}p{0.19\linewidth}p{0.45\linewidth}p{0.30\linewidth}@{}}
\toprule
\textbf{Threat} & \textbf{SAB Control} & \textbf{Residual Risk} \\
\midrule
Malicious proposer agent & Decouples proposal from execution; requires contract $C$ and signed certificate $\Omega$. & Bad proposals may consume validator and operator review capacity. \\
Compromised validator & Enforces validator diversity and critical archetype vetoes; signature quarantine. & A compromised validator can approve unsafe actions until quarantined. \\
Correlated validator failure & Diversity rules across model architectures and prompt structures. & Shared latent training biases may still fail to catch novel semantic errors. \\
Stale/manipulated evidence & Binds evidence digests $H(E)$; enforces freshness and provenance checks. & Upstream telemetry sources may report inaccurate state. \\
Forged/replayed certificate & Sovereign broker verifies $\sigma_{\mathrm{SAB}}$ with $pk_{\mathrm{SAB}}$, validity window $T_{\mathrm{valid}}$, and epoch $\rho_{\mathrm{rev}}$. & Compromise of airlock private key requires key rotation. \\
Policy downgrade attack & Binds signed, versioned policy releases $P_{\mathrm{ver}}$; enforces monotone routing. & Authorized policy updates may be overly permissive. \\
Human rubber-stamping & Structured evidence presentation; role separation; approval trends analysis. & Humans may still sign off negligent changes under operational pressure. \\
Emergency-mode abuse & Scoped emergency authority; short-lived certificates; mandatory post-hoc replay. & Social or operational abuse of emergency procedures. \\
Jurisdictional bypass & Location checks in policies; regional checks at the sovereign broker. & Incomplete or inaccurate asset tagging. \\
Autonomy creep & Autonomy level caps $\mathcal{L}(C)$; periodic ledger audits. & Slow normalization of deviance. \\
\bottomrule
\end{tabular}
\end{table*}

\subsection{Authority-Boundary Invariants}
The threat analysis yields four invariants for high-stakes agentic infrastructure:
\begin{itemize}
    \item \textbf{Proposal is not authority:} Agent proposals must compile to $C$ and be admitted by the airlock before execution.
    \item \textbf{Approval is evidence-bound:} Validator and human approvals must be cryptographically bound to $C$ and $H(E)$.
    \item \textbf{Execution is sovereign:} Infrastructure changes must execute through $ID_{\mathrm{exec}}$ at the broker, never via agent credentials.
    \item \textbf{Authority is revocable and replayable:} Certificates, validators, and autonomy levels must be revocable; decisions must be reconstructable from the ledger.
\end{itemize}
Under the stated trust assumptions, admitted autonomous actions are bounded by contract, evidence, certification path, broker verification, and revocation state, and remain auditable through the ledger.

\section{Discussion and Limitations}
\label{sec:discussion-limitations}

SAB limits when model reasoning can transition into system mutations. It does not guarantee that planning models are correct or aligned. The main operational questions concern review granularity, proportional assurance, sovereignty, composition, and residual risk.

\subsection{Operational Discussion}

\paragraph{Human Oversight and the Unit of Review.}
Traditional human-in-the-loop controls are prone to fatigue and rubber-stamping, especially when operators are presented with raw reasoning transcripts or unconstrained tool calls. SAB changes the unit of review: operators approve a normalized contract $C$, cryptographic evidence digest $H(E)$, consequence level $\mathcal{L}(C)$, and validator outputs. Approval is tied to a concrete contract, evidence snapshot, and policy context.

\paragraph{Proportional Assurance.}
Autonomy should be granted dynamically. The same agent may restart a stateless container (routed to \textsc{PolicyOnly}), but require SQA for rollbacks, human multi-signature for firewall modifications, and be prohibited from deleting databases. Consequence-aware routing directs latency and validator overhead to operations where mistakes have material operational, legal, or security outcomes.

\paragraph{Operational Sovereignty.}
SAB defines sovereignty operationally: the institution must maintain local control over policy evaluation, evidence retention, signing keys, revocation state, broker identity, and execution authority. External foundation models are restricted to proposal generation; they hold no production credentials and cannot bypass local certification and revocation.

\paragraph{Composition with OpenKedge Layers.}
In the OpenKedge stack, \openkedge provides bounded mutation interfaces; \sal separates planning from execution; \vai provides execution identity and evidence chains; \pdd validates generated code against schemas and protocol obligation invariants; and \sqa evaluates action-level semantics. SAB composes these layers into a system-level admission boundary.

\subsection{System Limitations}
SAB's security claims depend on several operational and design boundaries:
\begin{itemize}
    \item \textbf{Consequence Scoring Errors:} Consequence classification ($\mathcal{R}(C)$) is an operational classifier. It may underestimate risk when dependencies are hidden, or create bottlenecks by over-estimating risk and over-escalating low-risk tasks to human review.
    \item \textbf{Evidence Staleness and Incompleteness:} SAB relies on external telemetry and log sources. Stale or incomplete evidence can cause the airlock to certify actions under assumptions that no longer hold.
    \item \textbf{Validator and Policy Calibration:} SQA relies on validators whose reliability degrades after model, prompt, or workload changes. Additionally, SAB cannot correct policies that are incomplete, stale, or contradictory.
    \item \textbf{Prototype Generality:} Prototype results are workload-dependent and should not be interpreted as general performance guarantees.
    \item \textbf{Emergency Bypass Risk:} Outages may force operators to bypass normal assurance rules. While SAB records emergency overrides and enforces short validity windows, emergency access remains a critical threat vector.
    \item \textbf{Control-Plane Attack Surface:} SAB introduces high-value security targets (signing keys, the ledger, and the broker). Compromise of these trusted components invalidates all security properties.
\end{itemize}
These limitations motivate an incremental deployment strategy, starting with high-consequence domains (IAM, network policy, and deployments) and expanding as evidence pipelines and validator models mature.

\section{Related Work}
\label{sec:related-work}

We position SAB relative to systems security and safety mechanisms that partially address proposal admission, execution control, and auditability:

\paragraph{Runtime Assurance.}
Safety-monitor architectures (e.g., Simplex) supervise autonomous controllers and interpose fallback modes to prevent physical safety envelope violations~\cite{simplex-architecture,bloem-shield-synthesis,phan-neural-simplex}. The same interposition pattern can be applied to cloud and software control planes, where the monitored objects are discrete configuration changes, evidence freshness, and policy epochs.

\paragraph{Assurance Cases.}
Dependability and safety cases structure claims, assumptions, and evidence for pre-deployment system certification~\cite{rushby-assurance}. SAB brings a similar claim-evidence structure into runtime admission: approvals bind to specific contracts, evidence snapshots, and revocation state.

\paragraph{Distributed Consensus.}
Paxos, Raft, and PBFT protocols ensure ordered agreement on state transitions across replicated nodes~\cite{lamport-paxos,ongaro-raft,castro-pbft}. They do not verify whether those transitions are semantically correct or compliant with institutional rules. SAB treats authority as a separate admission question.

\paragraph{Access Control (IAM) and Policy Engines.}
Zero trust, IAM, and policy-as-code engines (e.g., OPA) enforce static permissions at API endpoints~\cite{rose-zerotrust,aws-iam,opa-rego,kubernetes-admission}. They are necessary inputs, but they authorize identities rather than validating autonomous proposals against live evidence. SAB adds proposal certification before broker execution.

\paragraph{AI Safety and Tool Governance.}
Agent safety tools (filters, sandboxes, critiques) reduce dangerous outputs from planning models~\cite{yao-react,inan-llamaguard}. They run as unprivileged utilities and cannot serve as control-plane security boundaries. In SAB, safety models can act as untrusted validators in the SQA mesh, but authority is released only by the airlock and broker.

\paragraph{Audit Logs and Evidence.}
Audit trails and transparency logs capture events for post-incident forensics~\cite{w3c-prov}. Certificate $\Omega$ binds pre-execution admission evidence, policy versions, and validator outputs; ledger-bound outcome record $O$ links subsequent observations to $\Omega$, making admission decisions replayable. Audit logs record executed events; the SAB certificate records the admission context that authorized execution.

\paragraph{Sovereign AI Governance.}
Sovereignty frameworks dictate local resource control and residency rules~\cite{nist-ai-rmf}. In SAB, those requirements appear as local policy evaluation, domestic validator meshes, and regional execution brokers, so external or mixed planning models cannot bypass local control boundaries.

Existing mechanisms authorize identities, replicate decisions, monitor runtime behavior, or record events. SAB combines these mechanisms into an admission boundary that decides when autonomous proposals may become execution authority.

\section{Conclusion}
\label{sec:conclusion}

Agentic infrastructure needs an authorization boundary between plausible model-generated proposals and privileged production mutations. The Sovereign Assurance Boundary (\sab) addresses that boundary through certificate-bound admission: agent outputs compile into execution contracts $C$; admission decisions bind those contracts to evidence digests $H(E)$, certification paths, policy versions, revocation epochs, and scoped broker identities; and execution proceeds only after the sovereign broker verifies the certificate $\Omega$.

The paper's core claim is intentionally narrow: SAB is not a replacement for IAM, formal verification, consensus, or human review, but a runtime admission path that composes those mechanisms so autonomous reasoning alone cannot become execution authority. The prototype and case studies show that the airlock--broker model can be implemented and measured across cloud operations, generated-code deployment, sovereign AI cloud settings, and regulated data workflows, while the limitations identify the remaining dependencies on evidence freshness, policy quality, validator calibration, and trusted control-plane components. SAB therefore treats delegated execution authority as certificate-bound, evidence-bound, revocable, and replayable, rather than as an ambient permission.

\bibliographystyle{unsrt}
\bibliography{refs}

\end{document}